\documentclass[12pt]{article}

\usepackage{hyperref}

\usepackage{graphicx}
\usepackage{amsmath}
\usepackage{amssymb}
\usepackage{slashed}
\usepackage{bm}
\usepackage{cite}
\usepackage{setspace}

\def\beq{\begin{equation}}
\def\eeq{\end{equation}}
\def\bea{\begin{eqnarray}}
\def\eea{\end{eqnarray}}
\def\bwt{\begin{widetext}}
\def\ewt{\end{widetext}}

\def \ll{\left|}
\def \rr{\right|}
\def \<{\left<}
\def \>{\right>}
\def \[{\left[}
\def \]{\right]}
\def \({\left(}
\def \){\right)}

\def \l.{\left.}
\def \r.{\right.}
\def \nn{\nonumber}
\def \nl{\nn\\}
\def \s{\sqrt{2}}
\def \st{\sqrt{3}}

\def \hf{\frac{1}{2}}
\def \Re{{\rm Re}}
\def \Im{{\rm Im}}

\def \la{\lambda}
\def \Jsi{{J/\psi}}
\def \ga{\gamma}
\def \Ga{\Gamma}
\def \eps{\varepsilon}
\def \ep{\epsilon}

\allowdisplaybreaks

\begin{document}

\begin{flushright}
UdeM-GPP-TH-14-236 \\
UMISS-HEP-2014-02 \\
\end{flushright}

\begin{center}
\bigskip
{\Large \bf \boldmath Probing New Physics in Higgs Couplings to
  Fermions \\ using an Angular Analysis} \\
\bigskip
\bigskip
{\large
Bhubanjyoti Bhattacharya $^{a,}$\footnote{bhujyo@lps.umontreal.ca},
Alakabha Datta $^{b,}$\footnote{datta@phy.olemiss.edu} \\
and David London $^{a,}$\footnote{london@lps.umontreal.ca}
}
\end{center}

\begin{flushleft}
~~~~~~~~~~~$a$: {\it Physique des Particules, Universit\'e
de Montr\'eal,}\\
~~~~~~~~~~~~~~~{\it C.P. 6128, succ. centre-ville, Montr\'eal, QC,
Canada H3C 3J7}\\
~~~~~~~~~~~$b$: {\it Department of Physics and Astronomy, 108 Lewis Hall, }\\
~~~~~~~~~~~~~~~{\it University of Mississippi, Oxford, MS 38677-1848, USA}
\end{flushleft}

\begin{center}
\bigskip (\today)
\vskip0.5cm {\Large Abstract\\} \vskip3truemm
\parbox[t]{\textwidth}{The standard-model Higgs boson couples to
  quarks through a parity-even scalar $Hq{\bar q}$ coupling. We show
  that the rare Higgs decay $H\to VZ$, where $V$ is a vector
  quarkonium state such as $\Jsi$ ($c{\bar c}$) or $\Upsilon(1S)$
  ($b{\bar b}$), can be used to search for the presence of a
  parity-odd pseudoscalar $Hq{\bar q}$ coupling. Since both $V$ and
  $Z$ can decay to a pair of charged leptons, this presents an
  experimentally-clean channel that can be observed at the
  high-luminosity LHC or a future hadron collider. The P-even and
  P-odd $Hq{\bar q}$ couplings can be measured by analyzing the
  angular distribution of the final-state leptons.}

\end{center}

\thispagestyle{empty}
\newpage
\setcounter{page}{1}
\baselineskip=14pt

\onehalfspacing

\section{Introduction}

There is no doubt that the scalar particle of mass 126 GeV recently
discovered at the LHC by the ATLAS and CMS Collaborations
\cite{Aad:2012tfa,Chatrchyan:2012ufa} is the Higgs boson. However, a
crucial question remains: is this the Higgs boson of the standard
model (SM), or do its properties indicate the presence of new physics
(NP)? To this end, there have been numerous studies of the Higgs
couplings to SM particles.

At tree level, the Higgs can decay to the $WW^*$ and $ZZ^*$ final
states. These decays have been measured, with the result that the
couplings of the scalar agree well with the theoretical predictions
for the SM Higgs boson \cite{Atlas:2014,CMS:2014}. The $ZZ^*$ state is
observed through its decay to four leptons, and as such it presents a
clean measurement channel. The $H\to 4l$ decay process has therefore
been used in several papers as a ``golden channel'' to look for NP
\cite{Modak:2013sb,Buchalla:2013mpa,Chen:2014gka,Beneke:2014sba}. At
the loop level, the Higgs can also decay to $\gamma\gamma$ and
$Z\gamma$. These decays can potentially probe the Higgs coupling to
the top quark \cite{Hgamgam}. The measured rate for $H\to \ga\ga$
agrees reasonably well with the SM prediction
\cite{Atlas:2014,CMS:2014}.

Although measuring the Higgs couplings in the bosonic decay channels
takes priority, directly measuring its couplings in fermionic modes is
also important. Indeed, one of the goals of the future LHC program is
to precisely measure the Higgs' couplings to all SM fermions. However,
this is challenging. Since the top quark is heavy, the favorable modes
for measuring the Higgs' coupling to top quarks involve Higgs
production in association with $t{\bar t}$, a single $t$, or a single
${\bar t}$ \cite{ggttH}. $H\to b{\bar b}$ and $H\to\tau^+\tau^-$
\cite{Atlas:2014} have been observed at the LHC, but a precise
measurement of the $H b{\bar b}$ and $H\tau^+\tau^-$ couplings will
require further investigation. A direct measurement of Higgs'
couplings to the first two generations of quarks is currently out of
reach of experiments, though a search for $H\to\mu^+\mu^-$ was
recently reported \cite{Hmumu}.

It seems clear that, in order to see evidence of NP in fermionic decay
modes of the Higgs, a significant improvement in sensitivity is
needed.  One potential way of improving the sensitivity to NP is to
study experimentally-clean modes such as those in which the final
state includes leptons. Such modes are rare but often free from
backgrounds.  One possibility is the decay $H\to l^+l^-\ga$, where $l$
represents an electron or a muon. This has been examined in
Ref.\ \cite{Bodwin:2013gca}.  Although this decay channel is extremely
rare due to the small SM $Hl^+ l^-$ coupling, it can receive a
significant contribution from the resonant production of a vector
quarkonium state $V$, in which the $V$ decays to an $l^+l^-$
pair. Examples of such a state are $\Jsi$ ($c{\bar c}$) or
$\Upsilon(1S)$ ($b{\bar b}$). In this case, the decay $H\to V\ga$
proceeds through the $Hq{\bar q}$ coupling. Higgs Yukawa couplings to
the first- and second-generation quarks can also be probed through
rare Higgs decays in which the final state consists of a QCD vector
meson and an electroweak gauge boson.  These channels were studied in
Ref.\ \cite{Kagan:2014ila} and deemed promising for observation at the
high-luminosity LHC and future hadron colliders.

The direct decay $H\to q{\bar q} \ga$, also known as the inverse
Wilczek process \cite{Wilczek:1977zn}, has been studied in
Refs.~\cite{Bodwin:2013 gca, Keung:1983ac}.
Ref.~\cite{Bodwin:2013gca} also considers the indirect decay process
$H \to \ga\ga^*$, in which the excited photon $\ga^*$ then decays to a
quarkonium state. The conclusion is that the interference between the
direct and indirect processes sufficiently enhances the rate that
$H\to V \ga$ can be observed at the high-luminosity LHC. In principle,
the study of this process will allow us to probe the NP properties of
the $Hq{\bar q}$ coupling.

In the SM, the $Hq{\bar q}$ coupling $c_S$ is purely scalar
(parity-even). In NP models, a parity-odd pseudoscalar coupling $c_P$
can be generated. The Higgs decays should therefore be studied with
the aim of detecting the presence of $c_P$. Now, the interference of
the SM scalar and NP pseudoscalar couplings will lead to P-odd
observables.  The examination of such observables will give
information about $c_P$, in particular whether it is nonzero.

However, this poses a problem. In $H\to V \ga$, the P-odd observable
is the triple product ${\vec q} \cdot ({\vec\varepsilon}_V^* \times
{\vec\varepsilon}_\gamma^*)$, where ${\vec q}$ is the difference
between the momenta of the $V$ and $\gamma$ in the rest frame of the
$H$, and ${\vec\varepsilon}_V^*$ and ${\vec\varepsilon}_\gamma^*$ are
the polarizations of $V$ and $\gamma$, respectively. But while ${\vec
  \varepsilon}_V^*$ can be measured by studying the momenta of the
leptons in $V \to l^+l^-$, ${\vec\varepsilon}_\gamma^*$ cannot be
measured since the photon does not decay. Thus, the process $H\to V
\ga$ cannot be used to obtain information about $c_P$.

Still, this also indicates how to resolve the problem. The photon in
$H\to V \ga$ must be replaced by a vector that does decay, so that its
polarization can be measured. The most obvious process is $H\to V Z$,
with $Z \to l^+l^-$. There are other possibilities, but this decay has
the largest rate and is easiest to observe experimentally, due to the
leptons in the final state. The process $H\to\Jsi Z$ was studied in
Ref.\ \cite{Isidori:2013cla}. (Note also that nonstandard $Hq{\bar q}$
couplings can enhance the decay rate \cite{Giudice:2008uua,
Englert:2014uua}.) In this Letter, we show how to test for the presence
of a nonzero $c_P$ by studying the angular distribution of $H\to V Z$.

In Sec.~\ref{Sec:II}, we introduce the P-odd $Hq{\bar q}$ coupling,
and examine how it can arise in NP models.  The matrix elements for
$H\to V\ga$ and $H\to VZ$ in terms of helicity amplitudes are
discussed in Sec.~\ref{Sec:III}. In Sec.~\ref{Sec:IV}, we show how to
separate and measure the P-even and P-odd $Hq{\bar q}$ coupling from
the helicity amplitudes using an angular analysis (full details are
given in the Appendix). We conclude in Sec.~\ref{Sec:V}.

\section{\boldmath $Hq{\bar q}$ Coupling}
\label{Sec:II}

We write the $Hq{\bar q}$ coupling in the following form:
\bea
{\cal L}_q &=& -\dfrac{m_q}{v}(c_S{\bar q}q + i c_P{\bar q}\gamma^5q)H ~.
\eea
Here $c_S$ and $c_P$ represent, respectively, the P-even scalar and
P-odd pseudoscalar couplings of the Higgs to a pair of quarks. In the
SM the coupling is purely scalar, so that $c_S = 1$ and $c_P = 0$.

Nonstandard Higgs couplings to fermions can arise in many theories
beyond the SM. These couplings can be modified compared to the SM
through mixing effects, when the SM Higgs boson mixes with other
scalars, or through NP corrections to the Higgs-fermion vertex
\cite{Englert:2014uua}. Higgs mixing effects are less interesting for
our purposes as they can be first probed in Higgs decays to gauge
bosons.

Modifications of the Higgs Yukawa couplings to fermions arising from
dimension-six operators in an effective field-theory framework have
been studied in several papers \cite{Buchmuller:1985jz,
  Grzadkowski:2010es,Delaunay:2013pja}. Below we focus on the up-type
quark sector, but a similar analysis holds for down-type quarks.  The
relevant operators are
\bea
    \mathcal{L}_{\rm EFT} = \lambda^u_{ij}  \bar{Q}_{i}\tilde H  U_{j}
	 +\frac{g^u_{ij}}{\Lambda^2}  \bar{Q}_{i}\tilde H U_{j} \left(H^\dagger H \right)
 + {\rm h.c.}
\label{LEFT}
\eea
Here the first term is the up-type Yukawa operator of the SM, while
the second term is a dimension-six operator suppressed by the NP scale
$\Lambda$. $Q_i$ and $U_i$ ($i=1,2,3$) are, respectively, the
left-handed quark doublets and right-handed quark singlets; $H$ is the
Higgs doublet, with $\tilde{H}=i\sigma_2H^*$.  $\lambda^u$ and $g^u$
are generic complex $3\times3$ matrices in flavor space. Setting the
Higgs field to its vacuum expectation value, we have
$H=(0,(v+h)/\sqrt{2})^T$. The mass and linear Higgs coupling matrices
are then
\bea
	M^u_{ij}
&=&	\frac{v}{\sqrt{2}}\left( \lambda^u_{ij} +  \frac{g^u_{ij}+g^{u*}_{ij}}{2}\frac{v^2}
{2\Lambda^2} \right)\,, \nn\\
	S^u_{ij}
&=&	\frac{1}{\sqrt{2}}\left( \lambda^u_{ij} + 3\frac{g^u_{ij}+g^{u*}_{ij}}{2} \frac{v^2}
{2\Lambda^2} \right)\,, \nn\\
A^u_{ij}
&=&	\frac{1}{\sqrt{2}}\left(3\frac{g^u_{ij}-g^{u*}_{ij}}{2} \frac{v^2}{2\Lambda^2} \right).\
\eea
Here $S^u_{ij}$ and $A^u_{ij}$ are the scalar ($\bar{Q}_{i}U_{j}$) and
pseudoscalar ($\bar{Q}_{i} \gamma_5 U_{j}$) couplings of the Higgs. In
general, as we go from the gauge basis to the mass basis by
diagonalizing $\lambda^u_{ij}$, flavor-changing neutral-current (FCNC)
couplings of the Higgs will be generated. As in
Ref.~\cite{Delaunay:2013pja}, we assume that $\lambda^u_{ij}$ and
$g^u_{ij}$ are aligned so as to avoid FCNC; such an assumption can be
justified in certain scenarios \cite{Datta:2008qn}. We further assume
that the only significant corrections occur for the charm-quark
couplings to the Higgs (or for the bottom-quark couplings to the Higgs
in the down-type quark sector).

In Ref.~\cite{Delaunay:2013pja} several different theoretical
frameworks are considered that can lead to an $Hq{\bar q}$ coupling
significantly larger than in the SM. These include a two-Higgs-doublet
model with minimal flavor violation (MFV) \cite{Trott:2010iz, Jung:2010ik,
Altmannshofer:2012ar, Dery:2013aba}, a general MFV \cite{Kagan:2009bn}
scenario with only one Higgs doublet, and composite models in which the
Higgs field is realized as a pseudo--Nambu--Goldstone boson (pNGB). In the
composite pNGB Higgs models, modifications of the Higgs couplings to
up-type quarks are parametrized by the effective Lagrangian in Eq.\ (\ref%
{LEFT}), with $\Lambda$ replaced by the global symmetry-breaking scale $f$,
the ``decay constant'' of the pNGB Higgs \cite{Manohar:1983md,
Giudice:2007fh}. Corrections to the $H c\bar c$ coupling are considered in
this framework in Ref.\ \cite{Delaunay:2013iia}, and it is found that, for
a fully-composite charm quark, a large enhancement of the coupling is
possible. There are interesting attempts to understand the small
light-quark masses in terms of suppressions from higher-dimensional
operators constructed from the Higgs field \cite{Giudice:2008uua,
Babu:1999me}. These models can lead to large modifications of the Higgs
couplings to the light fermions. Finally, we note that the P-odd
pseudoscalar coupling $c_P$ can be constrained from low-energy bounds on
electric dipole moments under certain assumptions, but the constraints for
the charm and bottom quark couplings are quite weak at present
\cite{Brod:2013cka}.

\section{\boldmath $H\to V_1V_2$: Amplitude}
\label{Sec:III}

\subsection{\boldmath $H\to V \gamma$}

\begin{figure}
\begin{center}
  \includegraphics[width=0.35\textwidth]{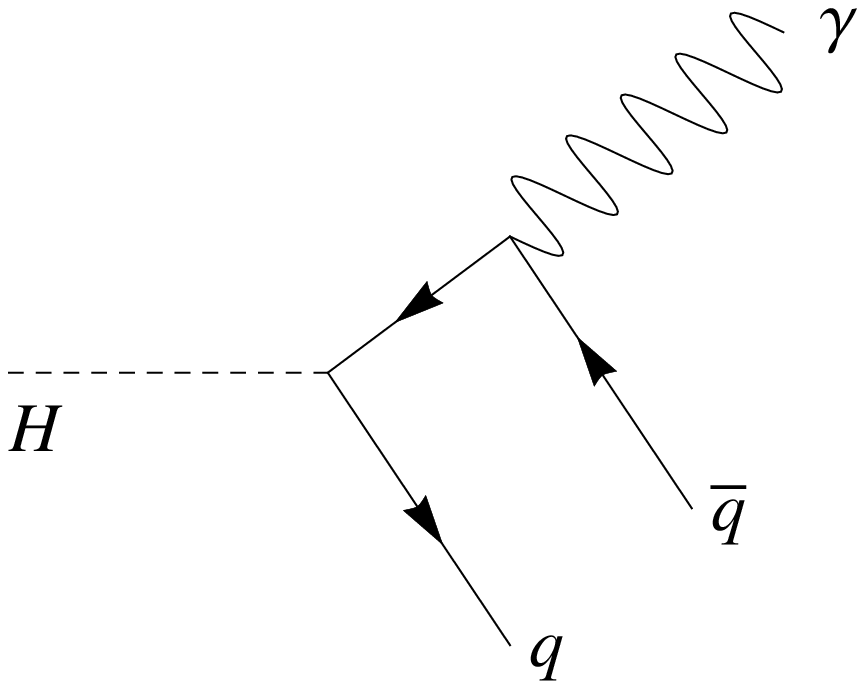}
  \includegraphics[width=0.35\textwidth]{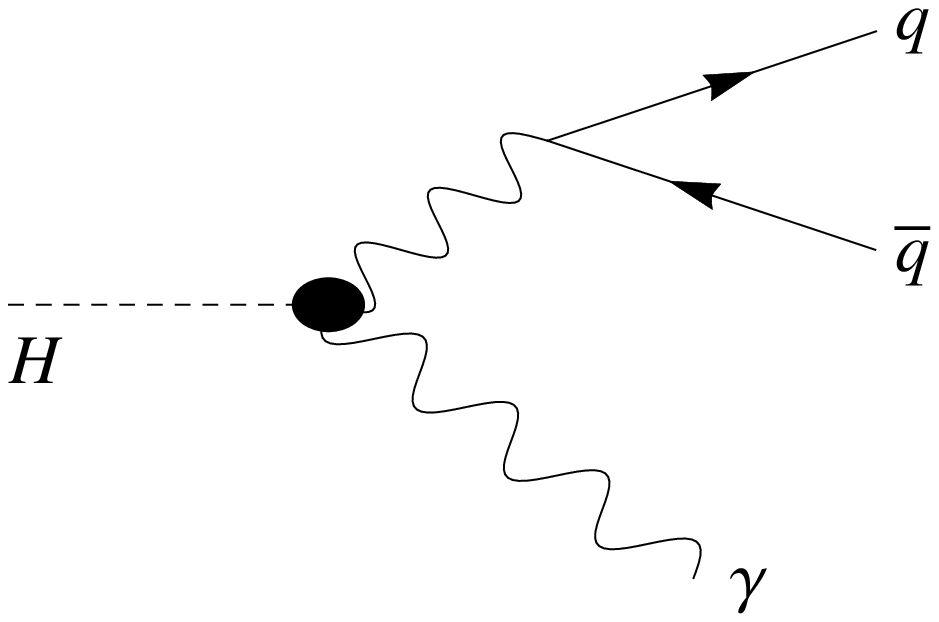}
\end{center}
\caption{Feynman diagrams for $H\to V\gamma$, where $V$ represents a
  $q{\bar q}$ state. The left-hand diagram involves the direct
  coupling of the Higgs to the quarks in $V$, while in the right-hand
  diagram the Higgs couples only indirectly to the quarks in $V$.
\label{fig:1}}
\end{figure}

In $H\to V \gamma$, the vector meson $V$ is a $q{\bar q}$ pair. The
Feynman diagrams for this decay are shown in Fig.~\ref{fig:1}.  At
tree level, the production of $V \gamma$ involves the $Hq{\bar q}$
coupling. This is shown in the left-hand diagram of Fig.~\ref{fig:1}.
At loop level, the vector can be produced from the decay of an
off-shell neutral gauge boson $\ga^*$ or $Z^*$. However, the
loop-level diagram shown on the right-hand side of Fig.~\ref{fig:1}
couples only indirectly to the quarks in $V$. As such, it does not give
rise to a P-odd term in the amplitude. We therefore focus primarily on
the tree-level diagram.

We begin by calculating the tree-level amplitude for $H\to q{\bar q}
\ga$. The final-state quark and antiquark then need to be dressed so
that they form the vector quarkonium state $V$, where the relative
motion between the $q$ and ${\bar q}$ within $V$ is small compared to
the large momentum of $V$ itself. This calculation can be done within
the framework of non-relativistic QCD (NRQCD)
\cite{Caswell:1985ui,Bodwin:1994jh} by expanding in powers of the
small relative velocity. For our purposes, we stick to the leading
order result in NRQCD where one neglects any relative motion between
the quark and the antiquark, so that the tree-level invariant matrix
element for $H\to V \gamma$ can be written as
\bea
{\cal M} &=& \dfrac{4\st e e_q\phi_0}{m^2_H - m^2_V}\(\dfrac{m_VG_F}{2\s}\)
^\hf\[c_S\{2(\eps^*_\ga\cdot p_V)(\eps^*_V\cdot k) - (m^2_H - m^2_V)(\eps^*
_\ga\cdot\eps^*_V)\}\r.~~ \nl
&&\hskip5truecm \l.-~2~c_P~\ep_{\mu\nu\rho\la}~\eps^{*\mu}_\ga~k^\nu~p^\rho
_V~\eps^{*\la}_V\]~,~~ \label{eq:VgaM}
\eea
where $\eps^*_{\ga(V)}$ is the polarization of the photon ($V$), $k$
and $p_V$ are the four-momenta of the photon and $V$, respectively,
and $\phi_0$ is the wave function of the $q{\bar q}$ state at zero
three-momentum. Since we neglect the relative motion of the quark and the
antiquark in $V$, $\phi_0$ can be considered real. Its magnitude can be
measured directly in experiments from the quarkonium decay to a pair of
leptons using the decay-rate formula
\bea
\Ga(V\to l^+l^-) &=& \dfrac{e^2_qe^4\phi^2_0}{\pi m^2_V}~.
\eea
Subleading NRQCD corrections give rise to a tiny phase in $\phi_0$
\cite{Bodwin:2013gca}, and also modify the coefficient of each term in
Eq.\ (\ref{eq:VgaM}). Ref.\ \cite{Bodwin:2014bpa} contains a detailed
discussion of the NRQCD corrections to the $H\to V\ga$ amplitude.  In
our first attempt to probe new physics in this decay, we neglect the
subleading contributions.

Equation (\ref{eq:VgaM}) can be written in a more familiar form by
going to the rest frame of the $V$. We can then define $\eps^{*L}_V
\equiv \vec\eps^*_V\cdot {\hat k}$ and ${\vec\eps}^{*T}_V \equiv
{\vec\eps}^*_V - \eps^{*L}_V~{\hat k}$. In the linear polarization
basis, also known as the transversity basis, we have
\bea
{\cal M} &=& H_\| \, {\vec\eps}^{*T}_V\cdot\vec\eps^*_\ga + i~H_\perp
{\hat k}\cdot({\vec\eps}^{*T}_V\times\vec\eps^*_\ga)~,~~
\label{MVgamma}
\eea
where
\bea
H_\| &=& 4\st e e_q\phi_0\(\dfrac{m_VG_F}{2\s}\)^\hf c_S ~,~~ \nn\\
H_\perp &=& 4\st e e_q\phi_0\(\dfrac{m_VG_F}{2\s}\)^\hf i~c_P~.~~
\label{Hperpdef}
\eea
Note that $H_\perp$ is proportional to $c_P$, so that it can arise
only if the pseudoscalar $Hq{\bar q}$ coupling is nonzero.

There are several things to notice about Eq.~(\ref{MVgamma}). First,
there is no term involving the longitudinal polarization. This is
because the final-state photon is on shell, and a massless particle
has no longitudinal polarization.  Second, the only P-odd observable
in $|{\cal M}|^2$ is the triple product (TP) ${\hat k} \cdot
({\vec\eps}^{*T}_V\times\vec \eps^*_\ga)$. It arises due to the
interference between the $H_\|$ and $H_\perp$ terms, and is
proportional to $c_S c_P$. Third, and most important, the measurement
of a nonzero value for the TP would indicate the presence of a NP
pseudoscalar $Hq{\bar q}$ coupling $c_P$. However, this requires
knowledge of the photon polarization $\eps^*_\ga$.  Unfortunately,
given that the photon does not decay, $\eps^*_\ga$ cannot be
determined. The upshot is that $H\to V \ga$ cannot be used to extract
information about $c_P$.

\subsection{\boldmath $H\to V Z$}

\begin{figure}
\begin{center}
  \includegraphics[width=0.3\textwidth]{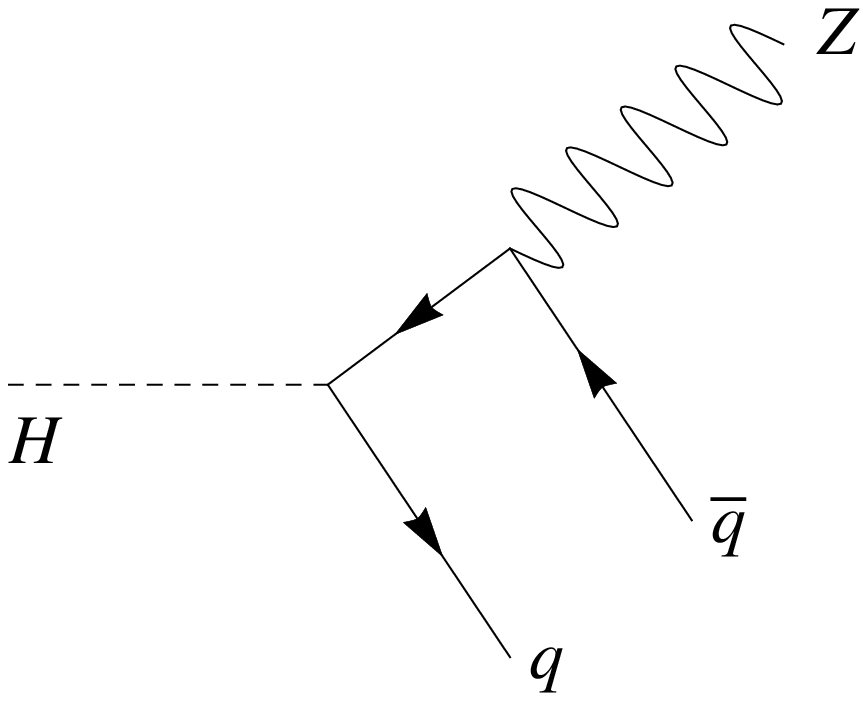}
  \includegraphics[width=0.3\textwidth]{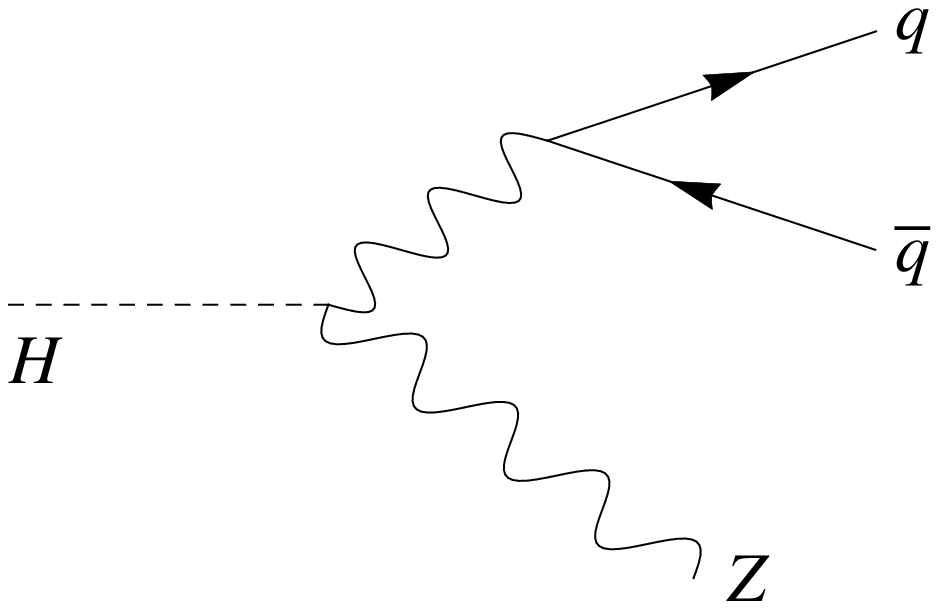}
  \includegraphics[width=0.3\textwidth]{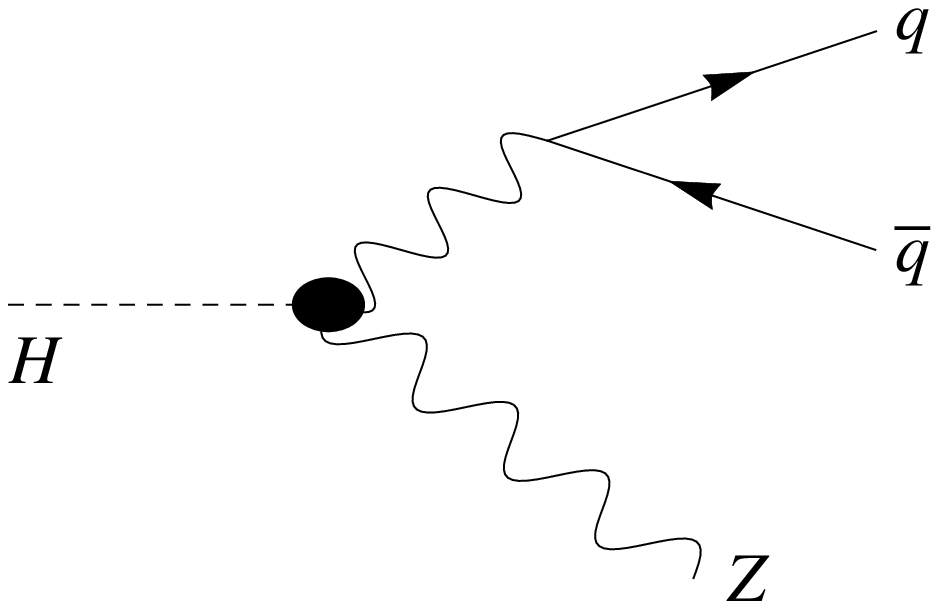}
\end{center}
\caption{Feynman diagrams for $H\to VZ$, where $V$ represents a
  $q{\bar q}$ state. The left-hand diagram involves the direct
  coupling of the Higgs to the quarks in $V$, while in the middle and
  right-hand diagrams the Higgs couples only indirectly to the quarks
  in $V$.
\label{fig:2}}
\end{figure}

The problem with $H\to V \gamma$ can be remedied by replacing the
photon with a vector that does decay, so that its polarization can be
measured.  This naturally leads us to examine $H\to V Z$. However,
unlike the photon, the $Z$ can couple to the Higgs at tree
level. Thus, there is an additional tree-level contribution to this
process, as shown in the middle diagram of Figure \ref{fig:2}.  Since
this diagram also contributes to the indirect coupling of the Higgs to
the quarks in $V$, just like the loop-level indirect-coupling diagram
on the right, it does not generate a P-odd term in the $H\to VZ$ decay
amplitude. In what follows we once again focus only on the
direct-coupling diagram.

As before, we can write down the leading-order NRQCD tree-level
invariant matrix element for the direct decay as
\bea
{\cal M} &=& \dfrac{\st c_V g\phi_0}{\cos\theta_W(m^2_H - m^2_V + m^2_Z)}
\(\dfrac{m_V G_F}{2\s}\)^\hf\[c_S\{2(\eps^*_Z\cdot p_V)(\eps^*_V\cdot p_Z) \r.\nl
&& \l.-~ (m^2_H - m^2_V - m^2_Z)(\eps^*_Z\cdot\eps^*_V)\} - 2~c_P~\ep_{\mu\nu
\rho\la}~\eps^{*\mu}_Z~p^\nu_Z~p^\rho_V~\eps^{*\la}_V\]~,~~
\label{eq:VZM}
\eea
where $c_V$ is the vector $Zq{\bar q}$ coupling in the SM. For up-type
quarks $c_V = 1 - (8/3)\sin^2\theta _W$, while for down-type quarks
$c_V = 1 + (4/3)\sin^2\theta _W$. Note that there is also an axial-vector
$Zq{\bar q}$ coupling. However, its contribution to the matrix element for
$H\to VZ$ vanishes to leading order in NRQCD.

Once again, in the rest frame of the $V$, Eq.~(\ref{eq:VZM}) takes a
more familiar form. Let ${\hat k}$ represent the direction of the $Z$
in this frame. With respect to ${\hat k}$ we can now define
longitudinal and transverse components of both the $V$ and $Z$
polarizations. In the linear polarization (transversity) basis, we
have
\bea
{\cal M} &=& H_0 \, {\vec\eps}^{*L}_V \cdot {\vec\eps}^{*L}_Z + H_\| \, {\vec\eps}^{*T}_V\cdot{\vec
\eps}^{*T}_Z + i~H_\perp {\hat k} \cdot ({\vec\eps}^{*T}_V\times{\vec\eps}^{*T}_Z)~,~~
\label{MVZ}
\eea
where
\bea
H_0 &=& \dfrac{\st c_V c_S g\phi_0}{\cos\theta_W}\(\dfrac{m_VG_F}{2\s}\)^\hf \dfrac{4m^2_V
m^2_Z}{(m^2_H - m^2_V - m^2_Z)(m^2_H - m^2_V + m^2_Z)}~,~~ \nn\\
H_\| &=& \dfrac{\st c_V c_S g\phi_0}{\cos\theta_W}\(\dfrac{m_VG_F}{2\s}\)^\hf\dfrac{m^2_H -
m^2_V - m^2_Z}{m^2_H - m^2_V + m^2_Z} ~,~~ \nn\\
H_\perp &=& i\dfrac{\st c_V c_P g\phi_0}{\cos\theta_W}\(\dfrac{m_VG_F}{2\s}\)^\hf\dfrac{
\Delta_V}{m^2_H - m^2_V + m^2_Z}~,~~ \nn\\
\Delta_V &=& \sqrt{(m^2_H - (m_V + m_Z)^2)(m^2_H - (m_V - m_Z)^2)}~.~~
\eea
As in Eq.~(\ref{Hperpdef}), $H_\perp$ is nonzero only if $c_P$ is nonzero.

As was the case for $H \to V \gamma$, we can see from Eq.~(\ref{MVZ})
that there is a P-odd TP ${\hat k} \cdot ({\vec\eps}^{*T}
_V\times{\vec\eps}^{*T}_Z)$ in $|{\cal M}|^2$ due to the interference
of the $H_\perp$ term with the $H_0$ or $H_\|$ terms. In this case, the
TP is measurable since ${\vec\eps}^{*T}_Z$ can be found by studying the
decay products of the $Z$. Since $H_\perp$ is proportional to $c_P$,
the nonzero measurement of the TP is a clear signal of a nonzero $c_P$.

Let us examine the helicity amplitudes in more detail. Consider the
decay $H\to\Jsi Z$, which can be used to probe the direct coupling of
the Higgs to $c{\bar c}$. Using $m_H = 125$ GeV, $m_Z = 91.2$ GeV, and
$m_\Jsi = 3.097$ GeV, we find
\bea
\dfrac{|H_0|}{|H_\||} &=& \dfrac{4 m^2_\Jsi m^2_Z}{(m^2_H - m^2_\Jsi -
m^2_Z)^2} \, = \, 6 \times 10^{-3} ~,~~ \nn\\
\dfrac{|H_\perp|}{|H_\||} &=& \dfrac{\Delta_\Jsi}{m^2_H - m^2_\Jsi - m^2_Z}
\dfrac{|c_P|}{|c_S|} \, \sim \, \dfrac{|c_P|}{|c_S|}~.~~
\eea
Thus, for the direct-coupling diagram, we see that the longitudinal
piece of the amplitude, $H_0$, is much smaller than a transverse
piece, $H_\|$. However, the magnitudes of the two transverse
components, $H_\|$ and $H_\perp$, can be comparable to one another if
$c_P$ and $c_S$ are of a similar size. In addition, the indirect decay
amplitude arising from the middle and right-hand diagrams in Fig.\ \ref
{fig:2} can contribute to $H_0$ and $H_\|$. These contributions have
been evaluated in \cite{Gonzalez-Alonso:2014rla,Gao:2014xlv}, and
their effect is generally to increase the magnitudes of both $H_0$ and
$H_\|$. However, these diagrams do not contribute to $H_\perp$ and
hence leave its structure unchanged.

The most complete study of $H\to V Z$ involves an angular analysis,
which permits the extraction of $H_0$, $H_\|$ and $H_\perp$. This is
discussed in the following section.

\section{\boldmath $H\to V Z$: Angular Analysis}
\label{Sec:IV}

In the Higgs rest frame, the $V$ and the $Z$ are back to back. Since
the Higgs is spinless, this decay distribution is isotropic. However,
the angular information obtained in $H\to VZ$ when the $V$ and
$Z$ each decay to a pair of leptons is sensitive to the helicity
amplitudes $H_0$, $H_\|$ and $H_\perp$. The analysis of the decay of a
scalar particle to a pair of vectors that subsequently decay to
leptons has been studied in the context of meson decays in
Refs.~\cite{Richman:1984gh, Kramer:1991xw, Dighe:1995pd,
  Bhattacharya:2013sga}.  Here we apply this technique to the decay
$H\to V Z$.

In the rest frame of the decaying Higgs, we choose our coordinates
such that the decay is along the $z$-axis. The subsequent decays
of the $V$ and the $Z$, each into a pair of leptons, can be
characterized in terms of three angles: the polar angles
$\theta_{V(Z)}$ corresponding to the $V(Z)\to l^+l^-$ decay axes in
the $V(Z)$ rest frames, and the azimuthal angle $\phi$ between the two
directions. Using the results of the Appendix, the differential decay
rate for $H\to VZ$ can be written as a function of $\phi$ as follows:
\bea
\dfrac{2\pi}{\Ga}\dfrac{d\Ga}{d\phi} &=& 1 + 4 \cos(2\phi) X + \hf \sin(2\phi) Y~,
\label{eq:fidis}
\eea
where
\beq
X = \dfrac{\ll H_\|\rr^2 - \ll H_\perp\rr^2}{\ll H_0\rr^2 + \ll H_\|\rr^2 +
\ll H_\perp\rr^2} ~~,~~~~
Y = \dfrac{\Im(H^{}_\|H^*_\perp)}{\ll H_0\rr^2 + \ll H_\|\rr^2 +
\ll H_\perp\rr^2} ~.
\eeq
Since $Y$ is linear in $H_\perp$, it is proportional to the P-odd
coupling $c_P$. Thus, the measurement of a nonzero $Y$ gives a clear
signal of NP in Higgs decays.

The advantage of performing an angular analysis with only $\phi$ is
that it does not require the high statistics needed to perform a
complete angular analysis. $Y$ can be simply extracted in experiments
as follows:
\bea
Y &=& \dfrac{\pi}{\Ga}\[\int\limits^{\pi/2}_0 \dfrac{d\Ga}{d\phi} d\phi
~- \int\limits^\pi_{\pi/2} \dfrac{d\Ga}{d\phi} d\phi ~+ \int\limits^{3\pi/2}
_\pi \dfrac{d\Ga}{d\phi} d\phi ~- \int\limits^{2\pi}_{3\pi/2} \dfrac{d\Ga}
{d\phi} d\phi\]~.
\eea
Similarly, one can extract $X$ using a different asymmetric integral
over $\phi$. Furthermore, since there are only two unknowns, $c_S$ and
$c_P$, the simultaneous measurement of X and Y allows one to obtain
both $c_S$ and $c_P$. Note that this holds even in the case that
$H_\|$ receives a significant contribution from the indirect coupling
of the Higgs to quarks via intermediate gauge bosons. Since
information about such couplings is available entirely from the Higgs
decay to gauge bosons, effectively $c_S$ and $c_P$ are still the only
unknown parameters.

Alternatively, one can use the full angular distribution for $H\to(l^+l^-)
^{}_V(l^+l^-)^{}_Z$ as a function of $\theta_V, \theta_Z$ and $\phi$ to
extract $H_0$, $H_\|$ and $H_\perp$. The derivation of the differential
decay rate for $H\to VZ$ is given in the Appendix. The result presented
there is model-independent, and simply describes a Higgs decay to a pair
of spin-one particles, each of which subsequently decays to a pair of
(massless) leptons. Combining the results in the Appendix with those in
Sec.~\ref{Sec:II}, we find
\bea
\dfrac{d\Ga}{d\cos\theta_Vd\cos\theta_Zd\phi} &=& |H^{}_0|^2~ W^{}_{00} ~+~
|H^{}_\||^2 ~W^{}_{\|\|} ~+~ |H^{}_\perp|^2 ~W^{}_{\perp\perp}~~~~~ \nl
&&\hskip-15truemm  ~+~ \Re\[H^{}_0H^*_\|\] W^{}_{0\|} ~+~ \Im\[H^{}_\|H^*_
\perp\] Y^{}_{\|\perp} ~+~ \Im\[H^{}_0H^*_\perp\]Y^{}_{0\perp}
~,~~~~~
\eea
where the $W's$ and $Y's$, which are functions of $\theta_V$,
$\theta_Z$ and $\phi$, are listed in Eq.~(\ref{eq:YL}). Asymmetric
angular integrations over $\theta_V$, $\theta_Z$ and $\phi$ can be
used to separate the coefficients of the angular functions. The
individual helicity amplitudes $H_0, H_\|,$ and $H_\perp$ can then be
obtained from a combined fit to these extracted coefficients.  Since
the three helicity amplitudes are functions of only two unknowns,
$c_S$ and $c_P$, one can solve for these unknowns, but with a certain
redundancy. This shows that the full angular analysis provides
additional cross checks for the validity of this formalism.

Finally, we note that the angular analysis presented in this section
is similar to that used to study $H \to Z l^+l^-$ in
Refs.\ \cite{Buchalla:2013mpa,Beneke:2014sba}.  These papers consider
the general distribution for $H \to Zl^+l^-$, which can in principle
include the contribution from $H \to VZ$, with the $V$ decaying to the
lepton pair.  However, in the SM the dominant contribution to this
decay comes at tree level from $H \to Z Z^*$, with the off-shell $Z^*$
decaying to the lepton pair. Because angular momentum is conserved in
both $H\to(l^+l^-) ^{}_V(l^+l^-)^{}_Z$ and
$H\to(l^+l^-)_Z(l^+l^-)_{Z^*}$, the expressions for the angular
distributions for both processes are similar \cite{Modak:2013sb}.  On
the other hand, while the study of $H\to(l^+l^-)_Z(l^+l^-)_{Z^*}$
sheds light on the coupling of the Higgs to gauge bosons, our primary
objective is to study the couplings of the Higgs to fermions.  As
explained earlier, a decay distribution of $H\to(l^+l^-)_V(l^+l^-)_Z$
that is asymmetric in the azimuthal angle $\phi$ can arise only due to
a P-odd direct coupling of the Higgs to the quarks in $V$.

\section{Conclusions}
\label{Sec:V}

Several new-physics scenarios suggest that the Higgs boson can couple
to quarks through a dimension-six operator which is odd under
parity. In this Letter, we discuss the consequences of such a P-odd
pseudoscalar coupling on the decay processes $H\to V\ga$ and $H\to
VZ$, in particular through triple-product (TP) correlations. Although
the pseudoscalar $Hq{\bar q}$ coupling gives rise to a TP in $H\to
V\ga$, it is not possible to retrieve information about the TP since
the photon polarization cannot be measured. We show that this problem
can be remedied by studying $H\to VZ$, in which both $V$ and $Z$ decay
to a pair of leptons. The dependence of the decay rate on the
azimuthal angle between the planes of leptons from the two decays can
be used to separate the P-even and P-odd couplings of the Higgs to the
quarks.

\section*{Acknowledgments}

This work was financially supported by the IPP (BB), by NSERC of Canada
(BB, DL), and  by the National Science Foundation (AD) under Grant No.\
NSF PHY-1068052. AD thanks Cedric Delaunay for useful discussion.

\section*{Appendix}

In the rest frame of the decaying Higgs, we choose our coordinates such
that the decay is along the $z$-axis. The amplitude for $H\to V Z$
depends on the helicities of the two vectors, and can be written as
\bea
A^{H\to VZ}_{\la_V,\la_Z} &=& D^{0*}_{0,\la_V - \la_Z}(0,0,0) H^{}_
{\la_V\la_Z} \nn\\
&=& \delta_{\la_V\la_Z} H^{}_{\la_V}~.
\label{eq:hvv}
\eea
Here $\la_X$ represents the helicity of the particle $X$,
$D^{J}_{M,M'}$ represents the Wigner $D$ functions, which are the
matrix elements of the rotation operator between eigenstates of
angular momentum, and $H$ represents the matrix elements for the Higgs
decay. Although $H$ depends on the helicities of both vectors,
angular-momentum conservation requires that the two vectors have the
same helicity. The helicity of a massive particle can take the values
0, $\pm1$.  Since the initial particle is spinless, its decay
amplitude is spherically symmetric, justifying our arbitrary choice of
coordinate axes.

We now allow the vector $V$ and the $Z$ to decay, each to an $l^+l^-$
pair.  The subsequent decay axis for the $V$ decay can be
characterized by a polar angle $\theta_V$ with respect to the
spin-quantization axis of the vector, chosen to be along the
$z$-axis. The decay axis for the $Z$ decay can be parametrized by a
second polar angle ($\theta_Z$) and an azimuthal angle ($\phi$). The
helicities of the final-state leptons can take the values $\pm
\hf$. However, the vector and axial-vector currents in the
electromagnetic and weak interactions of the SM require that the pair
of massless leptons have opposite helicities. Thus, without loss of
generality, the final-state helicity can be represented by
$\Delta\la_l = \la_{l^-} - \la _{l^+}$, which can take the values
$\pm1$. The amplitude for the decay $H\to(l^+l^-)^{}_V(l^+l^-)^{}_Z$
depends on the leptonic helicity differences for the two pairs of
final-state leptons ($\Delta\la_l$ and $\Delta\la'_l$), and can be
written as
\bea
A^{H\to(l^+l^-)^{}_{V}(l^+l^-)^{}_{Z}}_{\Delta\la_l,\Delta\la'_l} &=&
\sum\limits_{\la_V,\la_Z}\delta_{\la_V\la_Z}H^{}_{\la_V}D^{1*}_{\la_V,\Delta\la
_l}(0,\theta_V,0)V^{(V)}_{\Delta\la_l}D^{1*}_{\la_Z,\Delta\la'_l}(\phi,\theta_Z,
-\phi)V^{(Z)}_{\Delta\la'_l} \nn\\
&=& \sum\limits_\la e^{i(\la - \Delta\la'_l)\phi}d^{1}_{\la,\Delta\la_l}(\theta
_1)d^{1}_{\la,\Delta\la'_l}(\theta_Z)H^{}_{\la}V^{(V)}_{\Delta\la_l}V^{(Z)}_
{\Delta\la'_l}~.~~
\eea

The angular distribution for $H\to(l^+l^-)^{}_V(l^+l^-)^{}_Z$ can be
expressed as
\bea
\dfrac{d\Ga}{d\cos\theta_Vd\cos\theta_Zd\phi} &=& \sum\limits_{\Delta\la_l,
\Delta\la'_l}\ll A^{H\to(l^+l^-)^{}_V(l^+l^-)^{}_Z}_{\Delta\la_l,
\Delta\la'_l}\rr^2 \nn\\
&=&\sum\limits_{\la,\la'} e^{i(\la - \la')\phi} H^{}_\la H^*_{\la'} X^{(V)}_
{\la\la'}(\theta_V)X^{(Z)}_{\la\la'}(\theta_Z)~,~~ \label{eq:ad4l}
\eea
where
\bea
X^{(i)}_{\la\la'}(\theta_i) &=& \sum\limits_{\Delta\la_l} d^1_{\la,\Delta\la_l}
(\theta_i)d^1_{\la',\Delta\la_l}(\theta_i)|V^{(i)}_{\Delta\la_l}|^2~.~~
\eea
From the above, $X_{\la\la'}$ is symmetric under the exchange of $\la$
and $\la'$.  $\la$ and $\la'$ can each take the values 0 and
$\pm1$. Thus, we can write the six components as follows:
\bea
X^{(i)}_{++}(\theta_i) &=& \dfrac{1}{4}\[(1 + \cos^2\theta_i)\(|V^{(i)}_+|^2 +
|V^{(i)}_-|^2\) + 2\cos\theta_i\(|V^{(i)}_+|^2 - |V^{(i)}_-|^2\)\]~,~~ \nn\\
X^{(i)}_{--}(\theta_i) &=& \dfrac{1}{4}\[(1 + \cos^2\theta_i)\(|V^{(i)}_+|^2 +
|V^{(i)}_-|^2\) - 2\cos\theta_i\(|V^{(i)}_+|^2 - |V^{(i)}_-|^2\)\]~,~~ \nn\\
X^{(i)}_{00}(\theta_i) &=& \hf\sin^2\theta_i\(|V^{(i)}_+|^2
+ |V^{(i)}_-|^2\)~,~~ \nn\\
X^{(i)}_{+-}(\theta_i) &=& \dfrac{1}{4}\sin^2\theta_i\(|V^{(i)}_+|^2
+ |V^{(i)}_-|^2\)~,~~ \nn\\
X^{(i)}_{+0}(\theta_i) &=& \dfrac{\sin\theta_i}{2\s}\[\(|V^{(i)}_+|^2
- |V^{(i)}_-|^2\) + \cos\theta_i\(|V^{(i)}_+|^2 + |V^{(i)}_-|^2\)\]~,~~ \nn\\
X^{(i)}_{0-}(\theta_i) &=& \dfrac{\sin\theta_i}{2\s}\[\(|V^{(i)}_+|^2
- |V^{(i)}_-|^2\) - \cos\theta_i\(|V^{(i)}_+|^2 + |V^{(i)}_-|^2\)\]~.~~
\label{eq:Xs}
\eea

The above result is completely general. However, it simplifies when we
take into account certain properties of the $V$ and $Z$ decays. The
decay to $V\to l^+l^-$ is electromagnetic. Since the electromagnetic
interaction preserves parity, $|V^{(V)}_+| = |V^{(V)}_-|$. On the
other hand, the amplitude for $Z\to l^+l^-$ can be written as,
\bea
{\cal M} &=& \dfrac{g}{4\cos\theta_W}~\ep_\mu\[(c'_V + c'_A){\bar u}_R\ga^\mu
v_R + (c'_V - c'_A){\bar u}_L\ga^\mu v_L\]~,~~
\eea
where $c'_V = 4\sin^2\theta_W - 1$ and $c'_A = 1$. This implies that $|V^{(Z)}_-|
= \dfrac{c'_V - c'_A}{c'_V + c'_A}|V^{(Z)}_+|$. Using these, we can simplify our
earlier results. For the $V$ we find
\bea
X^{(V)}_{++}(\theta_V) &=& \dfrac{1}{4}(1 + \cos^2\theta_V)N_V~,~~ \nn\\
X^{(V)}_{--}(\theta_V) &=& \dfrac{1}{4}(1 + \cos^2\theta_V)N_V~,~~ \nn\\
X^{(V)}_{00}(\theta_V) &=& \hf\sin^2\theta_VN_V~,~~ \nn\\
X^{(V)}_{+-}(\theta_V) &=& \dfrac{1}{4}\sin^2\theta_VN_V~,~~ \nn\\
X^{(V)}_{+0}(\theta_V) &=& \dfrac{\sin2\theta_V}{4\s}N_V~,~~ \nn\\
X^{(V)}_{0-}(\theta_V) &=& -\dfrac{\sin2\theta_V}{4\s}N_V~,~~
\eea
where $N_V = |V^{(V)}_+|^2 + |V^{(V)}_-|^2$. For the $Z$ we find
\bea
X^{(Z)}_{++}(\theta_Z) &=& \dfrac{1}{4}\[(1 + \cos^2\theta_Z) + \dfrac{4 c'_v
c'_a}{c'^2_v + c'^2_a} \cos\theta_Z\]N_Z~,~~ \nn\\
X^{(Z)}_{--}(\theta_Z) &=& \dfrac{1}{4}\[(1 + \cos^2\theta_Z) - \dfrac{4 c'_v
c'_a}{c'^2_v + c'^2_a} \cos\theta_Z\]N_Z~,~~ \nn\\
X^{(Z)}_{00}(\theta_Z) &=& \hf\sin^2\theta_ZN_Z~,~~ \nn\\
X^{(Z)}_{+-}(\theta_Z) &=& \dfrac{1}{4}\sin^2\theta_ZN_Z~,~~ \nn\\
X^{(Z)}_{+0}(\theta_Z) &=& \dfrac{\sin\theta_Z}{2\s}\[\dfrac{2 c'_v c'_a}
{c'^2_v + c'^2_a} + \cos\theta_Z\]N_Z~,~~ \nn\\
X^{(Z)}_{0-}(\theta_Z) &=& \dfrac{\sin\theta_Z}{2\s}\[\dfrac{2 c'_v c'_a}
{c'^2_v + c'^2_a} - \cos\theta_Z\]N_Z~,~~
\eea
where $N_Z = |V^{(Z)}_+|^2 + |V^{(Z)}_-|^2$.

Thus the angular distribution for $H\to(l^+l^-)^{}_V(l^+l^-)^{}_Z$ of
Eq.~(\ref{eq:ad4l}) can be expressed as
\bea
\dfrac{d\Ga}{d\cos\theta_Vd\cos\theta_Zd\phi} &=& |H^{}_0|^2
\Omega^{}_{00}(\theta_V,\theta_Z) + |H^{}_+|^2 \Omega^{}_{++}(\theta_V,\theta_Z)
+ |H^{}_-|^2 \Omega^{}_{--}(\theta_V,\theta_Z)~~~~~\nl
&&\hskip-5truecm +~ \Re\[e^{2i\phi}H^{}_+H^*_-\]\Omega^{}_{+-}(\theta_V,\theta_Z)
+ \Re\[e^{i\phi} H^{}_+H^*_0\]\Omega^{}_{+0}(\theta_V,\theta_Z) + \Re\[e^{-i\phi}
H^{}_-H^*_0\]\Omega^{}_{0-}(\theta_V,\theta_Z)~,~~~~~
\label{eq:ad4l2}
\eea
where
\bea
\Omega^{}_{++}(\theta_V,\theta_Z) &=& \dfrac{1}{16}(1 + \cos^2\theta_V)\[(1 + \cos^2
\theta_Z) + \dfrac{4 c'_v c'_a}{c'^2_v + c'^2_a} \cos\theta_Z\] ~,~~ \nn\\
\Omega^{}_{--}(\theta_V,\theta_Z) &=& \dfrac{1}{16}(1 + \cos^2\theta_V)\[(1 + \cos^2
\theta_Z) - \dfrac{4 c'_v c'_a}{c'^2_v + c'^2_a} \cos\theta_Z\] ~,~~ \nn\\
\Omega^{}_{00}(\theta_V,\theta_Z) &=& \dfrac{1}{4}\sin^2\theta_V\sin^2\theta_Z~,~~ \nn\\
\Omega^{}_{+-}(\theta_V,\theta_Z) &=& \dfrac{1}{8}\sin^2\theta_V\sin^2\theta_Z~,~~ \nn\\
\Omega^{}_{+0}(\theta_V,\theta_Z) &=& \dfrac{\sin2\theta_V\sin\theta_Z}{8}\[\dfrac
{2 c'_v c'_a}{c'^2_v + c'^2_a} + \cos\theta_Z\]~,~~ \nn\\
\Omega^{}_{0-}(\theta_V,\theta_Z) &=& -\dfrac{\sin2\theta_V\sin\theta_Z}{8}\[\dfrac
{2 c'_v c'_a}{c'^2_v + c'^2_a} - \cos\theta_Z\]~.~~
\eea

Finally, it is also interesting to express all our results in the
transversity basis defined by $H^{}_\| = (H^{}_+ + H^{}_-)/\s$ and
$H^{}_\perp = (H^{}_+ - H^{}_-)/\s$.  In this basis, we can rewrite
Eq.~(\ref{eq:ad4l2}) as
\bea
\dfrac{d\Ga}{d\cos\theta_Vd\cos\theta_Zd\phi}
&=& |H^{}_0|^2 W^{}_{00}(\theta_V,\theta_Z,\phi) + |H^{}_\||^2 W^{}_{\|\|}(\theta_V,
\theta_Z,\phi) + |H^{}_\perp|^2 W^{}_{\perp\perp}(\theta_V,\theta_Z,\phi)~~~~~ \nl
&&\hskip-35truemm  +~ \Re\[H^{}_\|H^*_\perp\] W^{}_{\|\perp}(\theta_V,\theta_Z,\phi)
+ \Re\[H^{}_0H^*_\|\] W^{}_{0\|}(\theta_V,\theta_Z,\phi) + \Re\[H^{}_0H^*_\perp\]
W^{}_{0\perp}(\theta_V,\theta_Z,\phi) ~~~~~ \nl
&&\hskip-35truemm  +~ \Im\[H^{}_\|H^*_\perp\] Y^{}_{\|\perp}(\theta_V,\theta_Z,\phi)
+ \Im\[H^{}_0H^*_\|\] Y^{}_{0\|}(\theta_V,\theta_Z,\phi) + \Im\[H^{}_0H^*_\perp\]
Y^{}_{0\perp}(\theta_V,\theta_Z,\phi)~,~~~~~
\label{eq:ad4l3}
\eea
where
\bea
W^{}_{00}(\theta_V,\theta_Z,\phi) &=& \dfrac{1}{4}\sin^2\theta_V\sin^2\theta_Z~,~~
\label{eq:WF} \nn\\
W^{}_{\|\|}(\theta_V,\theta_Z,\phi) &=& \dfrac{1}{16}\[(1 + \cos^2\theta_V)(1 +
\cos^2\theta_Z) + \cos2\phi\sin^2\theta_V\sin^2\theta_Z\] ~,~~ \nn\\
W^{}_{\perp\perp}(\theta_V,\theta_Z,\phi) &=& \dfrac{1}{16}\[(1 + \cos^2\theta_V)
(1 + \cos^2\theta_Z) - \cos2\phi\sin^2\theta_V\sin^2\theta_Z\]~,~~ \nn\\
W^{}_{\|\perp}(\theta_V,\theta_Z,\phi) &=& \hf\dfrac{c'_vc'_a}{c'^2_v + c'^2_a}(1 +
\cos^2\theta_V)\cos\theta_Z~,~~ \nn\\
W^{}_{0\|}(\theta_V,\theta_Z,\phi) &=& \dfrac{1}{8\s}\sin2\theta_V\sin2\theta_Z\cos
\phi~,~~ \nn\\
W^{}_{0\perp}(\theta_V,\theta_Z,\phi) &=& \dfrac{1}{2\s}\dfrac{c'_vc'_a}{c'^2_v +
c'^2_a} \sin2\theta_V\sin\theta_Z\cos\phi~,~~ \nn\\
Y^{}_{\|\perp}(\theta_V,\theta_Z,\phi) &=& \dfrac{1}{8}\sin^2\theta_V\sin^2\theta_Z
\sin2\phi ~,~~ \nn\\
Y^{}_{0\|}(\theta_V,\theta_Z,\phi) &=& \dfrac{1}{2\s}\dfrac{c'_vc'_a}{c'^2_v + c'^2_a}
\sin2\theta_V\sin\theta_Z\sin\phi  ~,~~ \nn\\
Y^{}_{0\perp}(\theta_V,\theta_Z,\phi) &=& \dfrac{1}{8\s}\sin2\theta_V\sin2\theta_Z
\sin\phi ~.~~ \label{eq:YL}
\eea
Integrating Eq.~(\ref{eq:ad4l3}) over the polar angles $\theta_V$ and $\theta_Z$, it is
straightforward to obtain Eq.~(\ref{eq:fidis}).

\end{document}